\documentclass[11pt]{article}
\usepackage{iap2000,epsfig}
\usepackage{graphicx}
\usepackage[dvips]{color}
\def\be{\begin{equation}}
\def\ee{\end{equation}}
\def\bea{\begin{eqnarray}}
\def\eea{\end{eqnarray}}

\begin{document}
\vspace*{4cm}
\begin{center}
\title{Detection of Intracluster Gas Bulk Velocities in the Perseus \& Centaurus Clusters
\\[16pt]
}
\end{center}
\author{ Renato A. Dupke \& Joel N. Bregman }
\address{University of Michigan, Ann Arbor, MI 48109-1090}
\maketitle\abstracts{
We report the results of spatially resolved X-ray spectroscopy of 8 different {\sl ASCA}
pointings distributed symmetrically around the center of the Perseus cluster. The 
outer region of the intracluster gas is roughly isothermal, with temperature $\sim$ 6--7 
keV, and metal abundance $\sim$ 0.3 Solar. Spectral analysis of the central pointing
 is consistent with the presence of a cooling flow and a central metal abundance gradient.
  A significant velocity gradient is 
found along an axis highly discrepant with the major axis of the X-ray elongation. 
The radial velocity difference 
is found to be greater than 1000 km s$^{-1}Mpc^{-1}$ at the 90\% confidence level. Simultaneous 
fittings of GIS 2 \& 3 indicate that two symmetrically opposed regions have different radial velocities 
at the 95\% confidence level and the F-test rules out constant velocities for 
these regions at the 99\% level. Intrinsic short and long term variations of 
gain are unlikely (P $<$ 0.03) to explain the velocity discrepancies. 
We also report the preliminary results of a similar analysis carried out for the 
Centaurus cluster, where long-exposure SIS data is available. We also find a significant
velocity gradient near the central regions (3$^\prime$-8$^\prime$ of Centaurus. If attributed 
to bulk rotation the correspondent circular velocity is $\sim$1500$\pm$150~km s$^{-1}$ (at 90\% confidence).
The line of maximum velocity gradient in Centaurus is near-perpendicular to the infalling 
galaxy group associated with NGC 4709.}

\section{Introduction}

X-ray determination of the physical state of the intracluster gas provides a 
unique tool to probe the origin and evolution of clusters of galaxies.  {\sl ASCA} 
observations have shown significant spatial temperature variations in many clusters, 
indicating that clusters are currently evolving systems. This is consistent with the 
predictions of hierarchical cluster formation within CDM models, and had been 
suggested in pre-{\sl ASCA} times \cite{fit,u92}.

Although early models of galaxy clusters treated them as spherically symmetric 
virialized systems, 
recent X-ray and optical studies show that often clusters show substructures. 
Furthermore, if cold dark matter models (CDM) are correct, primordial small-scale density 
fluctuations are not erased and clusters are formed by infall/merging of smaller scale 
objects (bottom-up hierarchical scenario). The merger (infall) of sub-clumps creates 
shocks (associated with temperature substructure), bulk gas flows (associated with velocity 
substructure) and asymmetric distributions of velocity distribution of galaxies \cite{bird}. 
In order to understand the physical properties of clusters, their origin and evolution one has 
to take into account the degree and the physical scale of substructuring. 

The determination of complex temperature substructure in clusters is often  
interpreted to be related to shocks due to cluster merger at different stages. 
The link between temperature substructure and the merger stage is often done 
by comparison with hydrodynamical simulations. There is currently an enormous 
variety (different initial conditions, hydro-codes, impact parameters, matter 
components, etc.) of cluster formation/merger simulations in the literature 
\cite{ev,kw,rbl93,rbl97,sch,ptc,nfw,rlb97,emn,rick,tak98,tak99,tak00}, which provide simulated 
temperature maps that can be used to compare with the observations. Temperature 
maps provide important clues about the merging state of a cluster and the degree 
of substructuring, yet a comparison 
between temperature maps and numerical simulations is likely to show multiple 
configurations that may explain equally well what is observed, depending on the 
line of sight the observer chooses and also on the resolution of temperature maps, 
especially when one considers off-center mergers \cite{tak00,rick,rof}. This is 
because we are always looking at 2-D projections 
of the cluster, so additional information is needed descibe the cluster evolution 
stage accurately (e.g. gas velocity distribution).

A more straightforward way to determine the effects of a merger is to directly measure 
 intracluster gas velocities. Several simulations of cluster mergers \cite{rick,rbl93,tak99,rof} 
indicate residual gas velocities of a few thousand km~s$^{-1}$. To measure gas velocities one 
requires accurate determinations of spectral 
line centroids. The precision with which a line centroid can be measured, in velocity space, 
is $\sim 127~\Gamma_{eV}(E_{keV} N^{\frac{1}{2}})^{-1}$km~s$^{-1}$, where N is the number of photons in 
the line and $\Gamma_{eV}$ is the FWHM of the line, or if the line is narrower than 
the instrumental width, is the FWHM of the instrument, and E$_{keV}$ is the line energy.  
The energy resolution of the spectrometers on-board {\sl ASCA} 
 vary from 2\% (SIS) to 8\% (GIS)
at 5.9 keV. For a FWHM of 9000 km~$s^{-1}$ at 6.7 keV,  which is typical of early (first 3 years) 
SIS data at the FeK$\alpha$ line, to obtain a line centroid to a precision of 500 km/s, we need 
$\sim$ 60 line photons.  This same accuracy can be obtained with the GIS with $\sim$ 350 line photons.
Since these estimates become more uncertain if the rms of the continuum starts to compete with the line,
and, to compensate for the uncertainties introduced by {\sl ASCA} PSF, 
one would like to analyze clusters that have high metal abundance and
 several different long exposures observations 
from off-center regions surrounding the X-ray center.
Perseus (Abell 426) and Centaurus (Abell 3526) clusters match this criteria and, therefore, are 
well suited for the study of gas velocity distribution.

\section{The Perseus \& Centaurus Clusters}\label{subsec:prod}

The Perseus cluster (Abell 426) has been the subject of many studies since its discovery as 
an X-ray source \cite{fritz}. It is one of the closest (at an optical 
redshift of 0.0183), X-ray bright, rich cluster of galaxies. The cluster is 
elongated \cite{sny} and the ratio of its minor to major axis is 0.83 at radii greater than 
20$^\prime$. It has a cooling flow \cite{allen92,peres,ettori} with a mass deposition rate 
of about (3--5) $\times 10^{2}$ M$_{\odot}$ yr$^{-1}$.  The 
centroid of the cluster emission is offset by $\sim$ 2$^\prime$ to the east of NGC 1275 
\cite{sny,bran}. The average temperature \cite{eyles}  of 
the X-ray emitting gas is approximately 6.5 keV and the average 
abundance \cite{arnaud94} is 0.27 Solar in the central 1$^\circ$. The existence of an iron 
abundance gradient in Perseus was first suggested by 
Ulmer et al. \cite{u87} using data from SPARTAN 1. They found an iron abundance of $\sim$ 0.81
 Solar and a temperature of $\sim$ 4.16 keV within the central 5 $^\prime$ and an abundance 
 of $\sim$ 0.41 solar and a temperature of $\sim$ 7.1 keV in the outer regions 
(6 - 20$^\prime$). Further analyses have corroborated the existence of an abundance gradient 
\cite{pon,kowa,arnaud94,da}. 
Furthermore, the region where the abundance is enhanced is 
predominantly enriched by SN Ia ejecta \cite{da}, indicating that the 
cluster belongs to the class of clusters that present central ``chemical gradients'', 
such as A496 \cite{dw}. The presence of cooling flows, global abundance and chemical
gradients suggest that the cluster has not undergone strong mergers recently.

Centaurus (Abell 3526) is classified as Bautz-Morgan type I. It has an optical redshift 
of 0.0104. It has a small cooling flow and the estimated accretion mass rate is $<$ 
30-50 solar masses per year. Its X-ray emission is smooth, slightly elliptical in
shape and is strongly peaked on the cD galaxy NGC4696. The average temperature of the 
gas in the cluster is $\sim$ 3.5 keV \cite{thom,mat,allen94,mimu}. ASCA GIS \& SIS 
analysis of the central region (pointing) of Centaurus have 
shown evidence of a strong central abundance enhancement \cite{fuka} varying from Solar in the very 
central regions down to 0.3 Solar at $\sim$ 13$^\prime$,
also accompanied by a SN Ia ejecta dominance in the central region \cite{allen00}. 
Recently derived 2-D temperature maps of Centaurus (from GIS) suggest that a subgroup, which is 
centered on NGC 4709, may be merging with the cluster \cite{chu}. 

\section{Data Reduction \& Analysis}

{\sl ASCA} carries four large-area X-ray telescopes, each with its own detector: two Gas 
Imaging Spectrometers (GIS) and two Solid-State Imaging Spectrometers (SIS). Each 
GIS has a $50'$ diameter circular field of view and a usable energy range of 0.8--10 
keV; each SIS has a $22'$ square field of view and a usable energy range of 0.5--10 keV.
We selected data taken with high and medium bit rates, with cosmic ray rigidity
values $\ge$ 6 GeV/c, with elevation angles from the bright Earth of $\ge20^{\circ}$, 
and from the Earth's limb of $\ge5^{\circ}$ (GIS) or $10^{\circ}$ (SIS); we also excluded times when the 
satellite was affected by the South Atlantic Anomaly. Rise time rejection of particle 
events was performed on GIS data and
hot and flickering pixels were removed from SIS data. We estimated the background from blank 
sky files provided by the {\sl ASCA} 
Guest Observer Facility and removed bright point  sources for each instrument in all 
pointings. In the spectral fittings we used {\tt XSPEC} v10.0 \cite{arnaud96} software to analyze 
the GIS \& SIS spectra separately and simultaneously. The spectra were fitted 
using the {\tt mekal} thermal emission
models, which are based on the emissivity calculations of  Mewe \& Kaastra \cite{mewe85,mewe86,kaa},
with Fe L calculations by Liedahl et al. \cite{lied}. Abundances are 
measured relative to the solar photospheric values \cite{ag}, 
in which Fe/H=$4.68\times10^{-5}$ by number. Galactic photoelectric absorption was 
incorporated using the {\tt wabs} model \cite{mor}; Spectral channels 
were grouped to have at least 25 counts/channel. Energy ranges were restricted to 
0.8--9 keV for the GISs. 

Since there is a cooling flow at the center of Perseus we added a cooling 
flow component to the {\tt mekal} thermal emission model in the central pointing, 
to compare the temperature of the hot component in the center with that of the outer
pointings. For Centaurus the spectral analysis of all regions included a cooling flow
component. We tied the maximum temperature of the cooling flow to the 
temperature of the thermal component, and we fixed the minimum temperature at 0.1 keV. 
The abundances of the two spectral components ({\tt mekal} and {\tt cflow}) were tied
together. We also applied a single (but variable) global absorption to both spectral 
components. Since the cD galaxy of Abell 426 (NGC 1275) is an AGN, we also included a power
law component in the spectral fittings of the central pointing.

\subsection{Extraction Regions}

Eight individual pointings towards the direction of Perseus were analyzed 
in this work. The central pointing is the only one that includes the contaminating 
source (the center of the Perseus cluster). The other seven pointings are distributed
more or less symmetrically around the central pointing with an average distance of 
$\sim$ 40$^\prime$ from the center. Five of the pointings (the most recent ones) were 
consecutively observed in 1997 and are spaced in time typically by a day. The pointing
characteristics are shown in Figure 1. 

The SISs have a better spectral resolution (by a factor of 2--4) than the 
GISs. However, for Perseus, the analysis of SIS data for our pointings is severely limited by 
photon statistics (the GISs overall count rate is typically more than ten times that of the SISs),
making the SIS redshift determination very uncertain. This difference in count rate between 
GISs and SISs is due to the following reasons: 1) most of the pointings analyzed in 
this work were observed by the SIS in 2-CCD mode, which minimizes the energy resolution 
degradation due to the residual dark-current distribution and non-uniform charge 
transfer inefficiency effects\footnote{heasarc.gsfc.nasa.gov/docs/asca/newsletters/sis\_calibration5.html}; 
2) the X-ray center of Perseus is not in the detector's field of view for all outer pointings
analyzed and, therefore, most of the photons come from a 
specific direction (towards the cluster's center).  Furthermore, the GISs have a higher effective area at high 
energies. Therefore,  we use only GIS 2 \& 3 in the analysis of Perseus. 

We extracted spectra from a circular region with a radius of 20$^\prime$ for 
each pointing, centered at the detector's center (Figure 1). The best-fit values 
for temperature, abundance and redshift obtained from spectral 
fittings of GIS2 and GIS3 separately are consistent with those obtained through the 
spectral fittings of both GIS 2 \& 3 jointly. Therefore, we show here only the best-fit 
parameters of the simultaneous fittings. The resulting joint fits were very good with 
reduced $\chi^2_{\nu}\sim 1$ for all regions. 

\begin{figure}[t]
\vskip4.5truein
\includegraphics{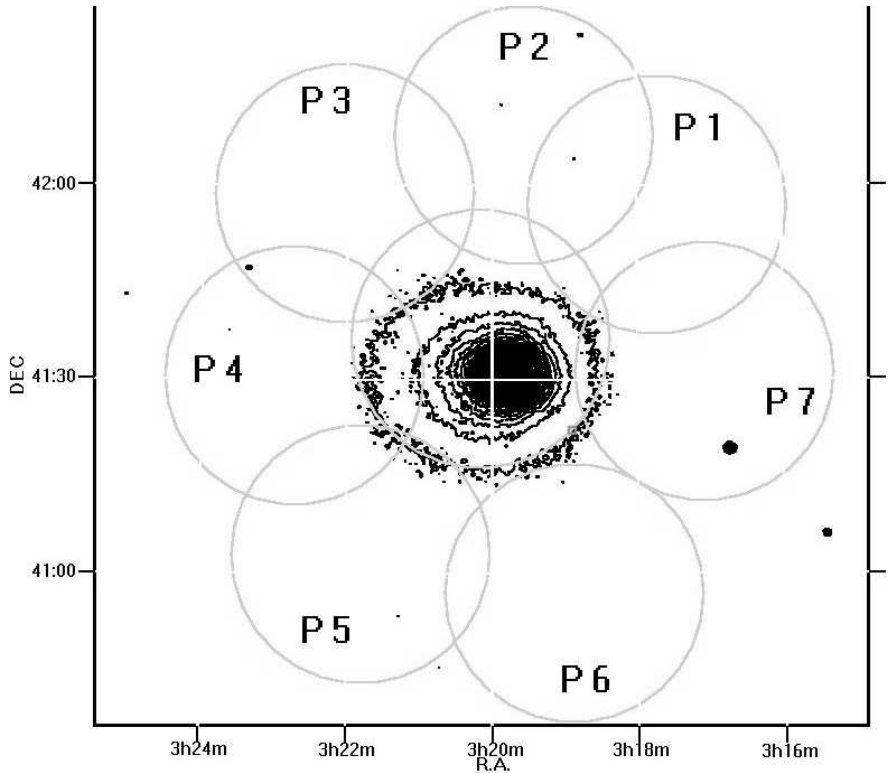}
\caption[]{Distribution of regions analyzed for different pointings analyzed in this work. PSPC surface 
brightness contours of Perseus are overlaid on the central pointing (P0).
The radius of each circular region is 20$^\prime$.\hfill
\label{fig:f1.ps}}
\end{figure}

We show here the preliminary results of spatially resolved spectroscopic analysis
of one long exposure (effective exposure $\sim$56 ksec and $\sim$68 ksec for GIS 
and SIS, respectively) taken on July of 1995 for the Centaurus cluster. The SIS observations were 
taken in 1-CCD mode using the best calibrated chips of each spectrometer 
(S0c1 and S1c3). We show here the results for a set of 9 different regions within the central pointing using
the SISs. The central region includes the core of Centaurus (P0) and the other regions are distributed
symmetrically around P0 and are centered 5$^\prime$ away. All regions considered have a radius of
3$^\prime$. These regions are illustrated in Figure 2.

\begin{figure}[t]
\vskip4.3truein
\includegraphics{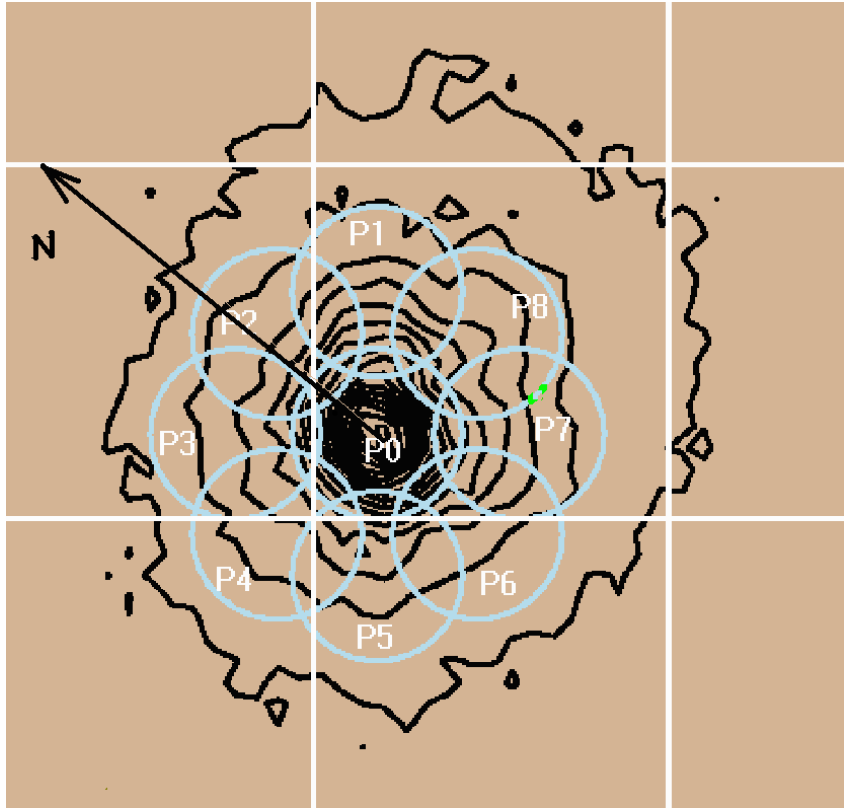}
\caption[]{Distribution of regions analyzed for the Centaurus cluster analyzed in this work. Surface 
brightness contours are overlaid. The North direction is indicated by the arrow.
The radius of each circular region is 3$^\prime$.\hfill
\label{fig:centmap-1.ps}}
\end{figure}

\section{Results}

\subsection{Perseus}

The best-fit values for temperature, abundance and redshift are
plotted in Figure 3 as a function of the azimuthal angle. Here, we define the azimuthal angle with
respect to the line that joins the centers of pointings P1(NW) and P5(SE), for convenience. 
For P1(NW) the azimuthal angle is set to zero.
 The indicated errors in Figure 3 are 
90\% confidence limits. It can be seen that Perseus appears to be roughly isothermal in
the outer regions  with an average temperature of 
$\sim$ 7 keV. The dashed and solid lines for the temperature plot in Figure 3 
indicate the 90\% confidence levels for the simultaneous GIS 2 \& 3 spectral fittings 
of the central pointing with and without the absorbed cooling flow component, 
respectively. The central best-fitting temperature for models that include a cooling 
flow component is not well constrained by the GISs and shows a best-fit value of 
6.85$\pm1$ keV, which is consistent with the observed temperatures in the outer 
regions. Since the GISs are not very sensitive to the absorbing column density the 
absolute values of the best-fitting temperatures may be artificially increased if 
the N$_{H}$ values are low.
 To test this effect we fixed N$_{H}$ at its putative Galactic value at the 
direction of each pointing (n$_{H}~\sim 1.6\times 10^{21} cm^{-2}$;
HEASARC NH Software) \cite{dick}. 
The best-fitting temperatures obtained this way have  significantly worse $\chi^{2}_{\nu}$ 
and are also 
plotted in Figure 3 (open circles). The best-fit temperatures obtained when 
N$_{H}$ is free to vary are lower by $\sim$ 1.5 keV than those obtained with free N$_{H}$. 
Although the azimuthal 
distribution of temperatures is consistent with isothermality, some 
pointings show mild, but significant, azimuthal variations.

The abundances observed in the outer parts of Perseus are generally lower than 
in the central region, which is consistent with observations by other authors \cite{u87,pon,kowa,arnaud94,da}. The 
abundance measured 
in the central pointing is 0.43$\pm$0.02 Solar and in the outer parts has an average value of 
$\sim$0.33 Solar. There is marginal evidence of azimuthal abundance variations. 
The best-fit abundances for each pointing are also displayed 
in Figure 3, where the solid lines represent the 90\% confidence limits for the 
abundance in the central pointing. The dash-dotted lines show the 90\% limits for 
the abundance measured within a 4$^\prime$ region of the center of Perseus for comparison \cite{da}.

The most important azimuthal distribution is that of redshifts. Two pointings 
show significant ($\ge$ 90\% confidence level) discrepant redshifts with respect to the 
best-fit redshift observed in the central pointing. These two redshift-discrepant 
pointings, P1(NW) and P5(SE), are on opposite sides of the cluster's center. 
P1(NW) shows a best-fit redshift of 0.014 (0.003-0.0185) and P5(SE) shows a significantly 
higher (95\% confidence level) redshift value of 0.042 (0.025-0.045). This redshift discrepancy 
is observed in both
GIS 2 and GIS 3 individually, although with lower statistical significance. 
This differences in redshifts imply a velocity difference $>$ 2000 
km s$^{-1}$ between these two pointings. The azimuthal distribution of redshifts is shown 
in Figure 3. In the spectral
fittings where the hydrogen column density is fixed at the Galactic value, the best-fit 
redshifts are typically lower than those obtained with N$_{H}$ free. However, the 
inferred differences between redshifts for different pointings are virtually unaltered. 
Therefore, the redshift discrepancies are not due to uncertainties related to the GIS
sensitivity to hydrogen column densities. 

\begin{figure}[t]
\vskip4.8truein
\includegraphics{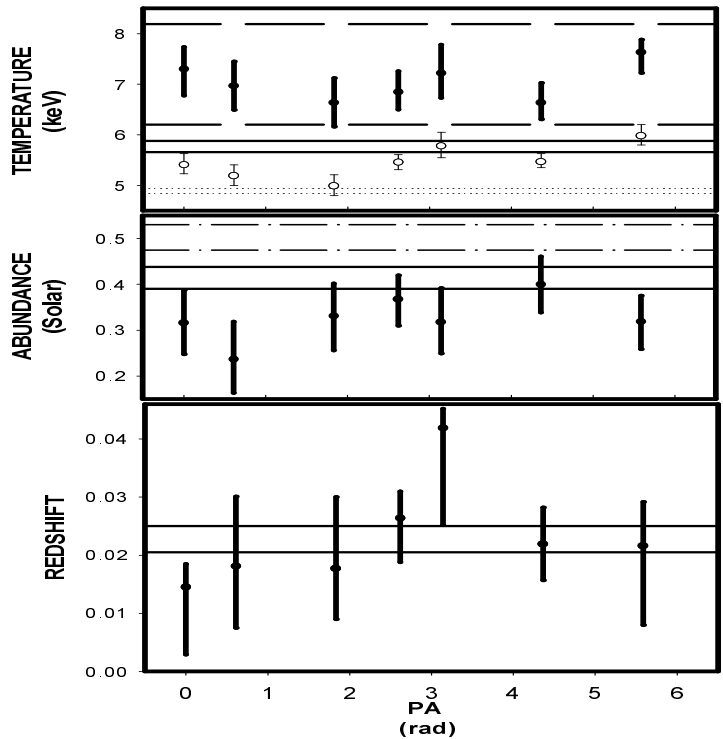}
\caption[]{Azimuthal distribution of Temperature (TOP), Abundance (MIDDLE), and Redshift (BOTTOM) as 
a function of the azimuthal angle (0--2$\pi$). First data point from the left for all plots corresponds
to P1, increasing to P7 (last). In the temperature plot solid and dashed lines represent the 
90\% confidence limits
for the central pointing (P0) without and with an extra cooling flow spectral component, 
respectively. Circles represent best-fit temperatures when N$_{H}$ is fixed at the Galactic 
value for each pointing. The dotted lines show the 90\% confidence limits
for the central pointing (P0) without the cooling flow spectral component and with N$_{H}$ fixed
at its Galactic value. 
In both the abundance and redshift plots solid lines show the 
90\% confidence limits for the central pointing. The dash-dotted lines show 
the best-fit abundance values for the central 4$^\prime$, for
comparison \cite{da}. Errors for all plots are 90\% confidence.\hfill
\label{fig:f2.ps}}
\end{figure}

The two redshift-discrepant pointings show no differences in the best-fit values of 
temperatures or abundances. To further test the significance of the velocity 
difference between P1 and P5 we simultaneously fit all four spectra (GIS 2 \& 3 for
each of the two pointings) and applied the F-test in the analysis of $\chi^2$ variations due to 
the change in the number of degrees of freedom. 
The difference in $\chi^2$ from the fits with locked and unlocked redshifts indicates 
that the redshifts are discrepant at the $99$\% level.
The inclusion of the power law component in the spectral fittings (representing any 
non-thermal emission from NGC 1275) in addition to the cooling flow component in
the central region (P0) does not change significantly the best-fit values of the redshifts 
measured without the power law component.

\subsection{GIS Gain Variations}

The original 
gain calibration of the GISs was mainly based on the built-in Fe-55 isotope source, 
attached to the edge of the field of view. The gain depends on the temperature of 
the phototube ($\sim$1\%/$^{\circ}$C), the position on the detector and time of 
observation. During the first several months in orbit the gain decreased by a few 
percent, and this trend has slowly disappeared. 
The intrinsic GIS gain is not only dependent on temperature but also on position 
(on the detector) due to non-uniformity in the phototube gain. The gain 
correction process involves a look-up table called the `gain map' \cite{tashi95},
which is also dependent on time and has been recalibrated using spectral lines 
observed during long ``day Earth'' and ``night Earth'' observations.
This allowed for more precise measurement of the azimuthal variation of gain across 
the detector, which is typically smaller than the radial gain variations.

The four gain corrections (short and long term gain variation, gain positional 
dependence and long term variation of the positional dependence), were carried 
out at GSFC in the standard processing (Ebisawa, private communication). 
Perseus observations analyzed in 
this work do have the standard gain corrections applied.
Since we still noticed small redshift fluctuations measured with GIS 2 and, 
especially, GIS 3 for the same 
region, we assume that, in spite of the standard gain correction, there are 
still residual gain variations and we also assume, conservatively, that the 
residual (post-standard processing) gain fluctuations magnitudes are 
on the same order as the fluctuations of 
the gain observed using the instrumental copper fluorescent line \cite{tashi99} at 8.048 keV. 
Since the gain fluctuations increase 
with time, we used the 1997 data as our standard of reference. The radial gain 
distribution shows that if one excludes the very outer region 
(r $\ge$ 22$^\prime$) and the very central region ($\le$ 2.5$^\prime$) from the spectral 
analysis the gain fluctuation around the mean is $\le$ 0.15\%, for both GIS 2 \& 3. For all pointings 
observed in this work we extract spectra from a circular region with a 20$^\prime$ 
radius.

However, the direction from which most photons are detected for pointings P1 and P5 are 
different, so we also need to estimate the azimuthal gain 
variations. We also use the gain map determined using the Cu-K line \cite{ide} at 8.048 keV. 
We compare the average gain values (excluding 
the outermost ring) of the region encompassing a 90$^{\circ}$ slice corresponding to 
the direction towards the real cluster's center (which is out of the field of 
view) for each instrument. Although the gain 
differences for the GIS 2 obtained this way imply a small gain variation ($\sim$ 0.12\%), for the 
GIS 3 the derived gain fluctuation is substantially larger ($\sim$ 0.37\%) than
that observed for the radial variations.

\subsection{Redshift Dependence on Gain}

In order to test the sensitivity of our observations to possible residual gain 
variations across the GISs we used Monte Carlo simulations. Supposing the redshift 
to be constant for the two discrepant regions, we generate fake spectra for both 
the GIS 2 \& 3 for the two pointings and compare the best-fit redshift differences.
We can then calculate the probability that we find the same redshift differences 
(or greater than) that we observe in the real pointings for GIS 2 \& 3.  We simulated 
1000 GIS 2 \& 3 spectra corresponding to each real observation using {\tt XSPEC} tool {\tt fakeit}. 

Each simulated spectrum was fitted in the same way as the real 
ones and the best-fit values of the redshift were recorded. We then selected the 
simulated spectra that had a redshift difference equal to or greater than 
that observed in the real pointings for both GIS 2 \& 3 (0.0208 for GIS2 and 
0.0302 for GIS3). In order to introduce the effects 
of GIS gain variations, we added a gain uncertainty to the best-fit 
values of redshift derived from fake spectra. We assume that the gain variations 
follow a Gaussian distribution with a standard deviation ($\sigma_{gain}$), 
which is different for each spectrometer, and a zero mean. This gain uncertainty 
is then summed to the best-fit redshifts obtained from the fake spectra before calculating the 
redshift differences between different pointings. To be conservative we assumed as 
our 1-$\sigma$ gain variations ($\sigma_{gain}$) for the GIS 2 \& 3 
the largest values of the two procedures described above, i.e. 0.15\% and 0.37\% for 
GIS 2 \& GIS 3, respectively.

The probability of observing the redshift differences in GIS 2 \& 3 that we measure 
for the real spectra in two pointings (P1 and P5), using the procedure described 
above is found to be 0.005. To illustrate how sensitive this value is to the assumed 
$\sigma_{gain}$ we varied the estimated $\sigma_{gain}$ and recalculated the 
probability of finding the redshift differences by chance. We find \cite{db1} that 
this probability is insensitive to gain variations up to a $\sigma_{gain}$ of $\sim$ 0.5\%. 
In a more realistic case we observe overall seven outer pointings and not only two. Therefore, 
we estimate the probability of finding the same redshift differences that we see 
in pointings P1 and P5 in seven observations using the same procedure to simulate 
spectra as described above. We also include a condition for alignment (slice encompassing 120$^\circ$). 
Even in this case, the velocity difference is still significant at a 97\% confidence level.

\subsection{Centaurus}

The temperature and abundance radial distributions in Centaurus are consistent with the presence of 
a cooling flow and a central abundance enhancement.
The temperature rises from a central value of 2.65$\pm$0.02 keV to an average value of $\sim$ 3.2 keV
within the central $\sim$ 5$^\prime$. The abundance declines from 1.27$\pm$0.04 Solar to an average
value of $\sim$0.96 Solar. There are significant ($\ge$90\% confidence) azimuthally symmetric temperature 
variations (Figure 4). The maximum temperature (3.6$\pm$0.11) is obtained for a position angle of 
$\sim$ 135$^\circ$ towards the direction of the high temperature ($\sim$ 4.5-5 keV) 
infalling subgroup associated with NGC4709 \cite{chu}. 

A significant velocity gradient is found between the central pointing (P0) and three 
other pointings P3(NE), P4(E) and P6(S). The velocity difference between 
these pointings was also noticed in SIS 0 \& 1 separately but with lower significance. To compensate 
for global gain variations between the different chips a gain correction of $\sim$ 1\% is 
applied to SIS 0 prior to performing simultaneous fittings. The maximum velocity difference found 
is $\sim (2.7\pm0.25)~\times~10^3$ km~s$^{-1}$ (errors are propagated 90\% confidence) between 
P3 and P6. The velocity differences with respect to the central pointing (P0) for both 
P3 and P6 are $\sim$ $\pm (1.38\pm0.15)~\times~10^3$ km~s$^{-1}$. If the velocity discrepant
pointings are interpreted as bulk rotation the apparent rotation axis orientation is
similar to the projected direction of the infalling group. 

The velocity differences found are significantly larger than what might be 
expected from residual (post-processing) gain variations within the two chips 
used. Ni fluorescence lines in the SIS background gain calibration for 1-CCD 
mode brings the uncertainties within these two chips to $\sim$ 0.27\% 
(heasarc.gsfc.nasa.gov/docs/asca/4ccd.html). A detailed analysis of 
the gain variation effects on the results described here is shown elsewhere \cite{db2}. 
 
\begin{figure}[t]
\vskip4.8truein
\includegraphics{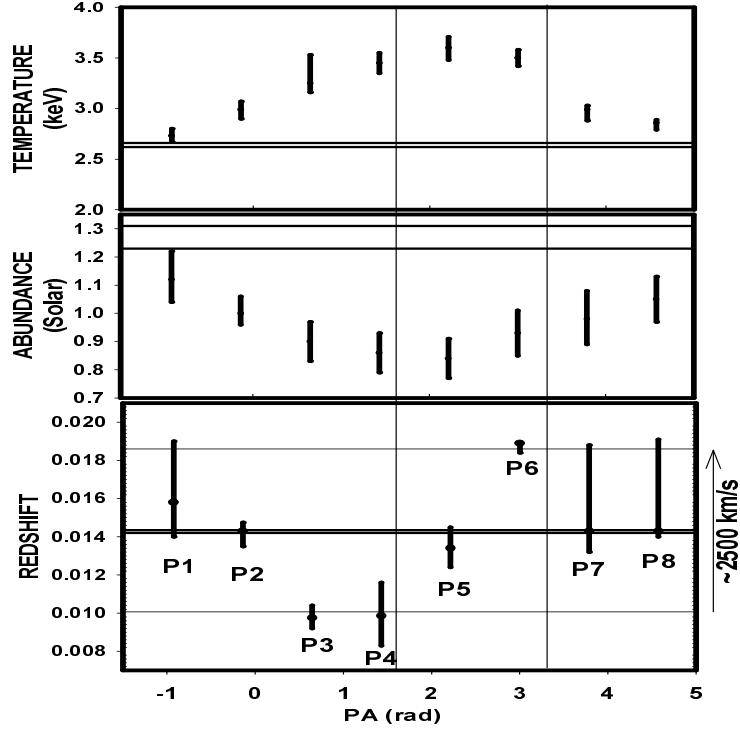}
\caption[]{Azimuthal distribution of Temperature (TOP), Abundance (MIDDLE), and Redshift (BOTTOM) as 
a function of the position angle (radians) for the Centaurus cluster. First data point for all 
plots corresponds to P1, increasing to P8 (last). For all plots the thick horizontal solid lines 
represent the 90\% confidence limits
for the central pointing (P0). Solid vertical lines show the approximate direction of the hot 
component associated with NGC 4709 \cite{chu}. The thin horizontal lines show the approximate velocity
difference equivalent to 2500 km~s$^{-1}$. Errors for all plots are 90\% confidence.\hfill
\label{fig:z_s01_cf.ps}}
\end{figure}

\section{Discussion}

The spectral analysis of {\sl ASCA} data carried out in this work indicates the 
presence of a significant velocity gradient in the ICM of the Perseus \& Centaurus clusters. 
In Perseus two regions symmetrical with respect to the cluster center show 
discrepant redshifts not just with respect to each other but also with respect to the central region.
This velocity difference is unlikely to be 
attributed purely to gain fluctuations and suggests the existence of large-scale bulk motions of 
the intracluster gas. The two symmetrically opposed discrepant regions have velocity differences of $\sim$~-3000 
km~s$^{-1}$ (or $\le$ -~600 km~s$^{-1}$ at the 90\% confidence level) for P1 and 
$\sim$ +5000 km~s$^{-1}$ (or $\ge$ +~60 km~s$^{-1}$ at the 90\% confidence level) 
for P5 with respect to P0. There are no observed temperature or abundance differences 
between the two pointings with discrepant redshifts (P1 \& P5).
The velocities measured are consistent with large scale gas rotation with a correspondent
circular velocity of $\sim$ 4100 $^{+2200}_{-3100}$ km s$^{-1}$ (90\% confidence) \cite{db1}. This  
implies a large angular momentum for the ICM
and that a significant fraction of the gas energy may be kinetic.  

In Centaurus, where analysis of SIS data allow more precise velocity measurements, we also
find evidence for significant velocity differences consistent with bulk rotation with 
circular velocity of $\sim$ 1500$\pm$150~km s$^{-1}$ (at the 90\% confidence) at 5$^\prime$ from the 
cluster center. There is also evidence for an azimuthal temperature gradient in 
Centaurus, where the temperature grows from 2.73$\pm$0.06 keV in the 
northwestern direction to 3.6$\pm$0.11 in the Southeastern region, towards the 
apparently merging galaxy group centered in NGC 4709. 

The best candidate for generating this large angular momentum in both clusters is off-center 
mergers. In off-center mergers, up to $\sim$ 30\% of the total merger energy may be kinetic
(can be transferred to rotation) \cite{ptc}. Off-center merger simulations \cite{rick,tak99,tak00,rof} 
often produce residual intracluster gas rotation with velocities 
of a few thousand km~s$^{-1}$.
Additional evidence for merging comes from the observed offset between the 
optical center and the X-ray center (Perseus) \cite{bran,sny,u92}; 
the radial change in X-ray isophotal orientation (isophotal twist) \cite{mohr}; 
the asymmetric galaxy morphological distribution, with preferential eastward direction in the 
distribution of E+S0 types (Perseus) \cite{bru}; the bimodal velocity distribution coincident 
with temperature substructure (Centaurus) \cite{chu,lucey}. 
However, simulations also indicate other observable consequences of mergers that can, in 
principle, be cross-analyzed with the velocity maps to test the robustness of the merger 
scenario. One of the features predicted by off-center cluster-cluster mergers is a 
strong negative radial temperature gradient (core heating) ($\ge$ 2 keV/Mpc) 
for most of the merger life-time, even when the angle of view is not favored, e.g. 
along the collision axis \cite{tak00,rick}. 
In our case we do not observe a negative temperature gradient at all, actually we 
observe a positive temperature gradient in both clusters due to the cooling flow. 
The mere fact that the cooling flow is present makes the off-center merger 
explanation more uncertain, since it 
has been suggested \cite{esf,rbl93} that a merger would disrupt any pre-existing cooling flows. However,
recent simulations of head-on cluster mergers indicate that cooling flows can survive
mergers depending on the produced ram-pressure of the gas in the infalling cluster \cite{gomez}. 
Even if cooling flows are disrupted by a merger they can be reestablished quickly if the cooling time
of the primary pre-merger component is small \cite{gomez}. The fact that the major 
axis of the X-ray elongation is relatively close to the apparent 
rotation axis is another difficulty posed to the off-merger explanation. In most cases 
simulations show that the 
isodensity contours are elongated perpendicularly to rotation axis, except in some 
short-lived merger stages viewed from specific directions \cite{tak00}. 

The velocity and temperature distributions of Centaurus allow for the possibility that we 
are seeing a pre-merger stage with the (NGC 4709) sub-group infalling with a significant
velocity component 
on the line of sight and on the E-W directions. In the case of Perseus, however, 
a pre-merger scenario does not fit the observations easily. This is, in part, because 
we do not observe X-ray surface brightness 
enhancement at the direction of the redshift-discrepant regions and also because there are 
no clear temperature substructures associated with the regions of discrepant velocities
\footnote{We do observe a mild enhancement in surface brightness 
towards the East region of Perseus, coinciding with our P4 pointing. This enhancement was noticed 
previously \cite{schw,ettori} with better significance, 
and has been interpreted as evidence for merging. Although there is 
no clear optical evidence of a sub-group towards the direction of P5 that corroborates 
this scenario, the region of surface brightness enhancement (P4) seems to be associated 
with a higher fraction of early-type galaxies \cite{bru}.}. 
Given that in some off-center merger simulations high rotational gas bulk
velocities can be maintained for $\ge$ 3 
crossing times \cite{rick} and that cooling flow survival to mergers is more likely 
to happen in off-center mergers \cite{gomez} the results are consistent with a large 
off-center merger event in Perseus $\ge$ 4 Gyr ago (assuming a cluster mass \cite{ettori} of 
5$\times$10$^{14}$ M$_{\odot}$ at a radius of 1 Mpc and 
that the rotating gas at $\sim$ 7 KeV is gravitationally bound). 

Velocity measurements of intracluster gas with 
{\sl Chandra} and, especially, {\sl XMM-Newton}
satellites will be able determine ICM velocities in the central regions more precisely, thus 
providing information on the gas velocity curve, which will strongly constraint 
cluster-cluster merger models, or suggest alternatives for generating the large 
angular momentum observed.

\section*{Acknowledgments}
We would like to thank, J. Arabadjis, G. Bernstein, M. Sulkanen, M. Ulmer 
 and M. Takizawa for helpful discussions, 
J. Irwin and P. Fischer for  helpful discussions and suggestions. The authors 
would particularly like 
to thank Dr. K. Ebisawa for providing information about {\sl ASCA} GIS gain calibrations that
was crucial to this work. We acknowledge support 
from NASA Grant NAG 5-3247. This research made use of the HEASARC {\sl ASCA} database and NED.

\end{document}